\documentclass[12pt]{article}
\usepackage{enumerate}
\usepackage{amsfonts}
\usepackage{amsmath}
\usepackage{amssymb}
\usepackage{amsthm}

\def\H{\mathcal{H}}

\def\P{\mathcal{P}}
\def\S{\mathfrak{S}}

\def\T{\mathfrak{T}}
\def\B{\mathfrak{B}}

\newcommand{\supp}{\mathrm{supp}}
\newcommand{\rank}{\mathrm{rank}}
\newcommand{\id}{\mathrm{Id}}
\newcommand{\Tr}{\mathrm{Tr}}

\newcounter{defin}  \newcounter{lemma}  \newcounter{theorem}
\newcounter{property} \newcounter{corol}  \newcounter{remark} \newcounter{example}

\newenvironment{lemma}{\par\refstepcounter{lemma}
     \textbf{Lemma \thelemma.} }{\rm\par}
\newenvironment{theorem}{\par\refstepcounter{theorem}
     \textbf{Theorem \thetheorem.}\ }{\rm\par}
\newenvironment{property}{\par\refstepcounter{property}
     \textbf{Proposition \theproperty.}\ }{\rm\par}
\newenvironment{corollary}{\par\refstepcounter{corol}
     \textbf{Corollary \thecorol.} }{\rm\par}
\newenvironment{definition}{\par\refstepcounter{defin}
     \textbf{Definition \thedefin.}\ }{\rm\par}
\newenvironment{remark}{\par\refstepcounter{remark}
     \textbf{Remark \theremark.}}{\rm\par}
\newenvironment{example}{\par\refstepcounter{example}
     \textbf{Example \theexample.}}{\rm\par}

\begin{document}
\title{On equalities in two entropic inequalities\footnote{This work was
partially supported by the
program ``Mathematical control theory and dynamical systems'' of RAS and by the RFBR grants 12-01-00319a and 13-01-00295a.}}
\author{M.E. Shirokov\\
Steklov Mathematical Institute, RAS, Moscow\\
msh@mi.ras.ru}
\date{}
\maketitle

\begin{abstract}
A simple criterion for local equality between the constrained Holevo
capacity and the quantum mutual information of a quantum channel is
obtained. It implies that the set of all states for
which this equality holds is determined by the kernel
of the channel (as a linear map).

Applications to Bosonic Gaussian channels are considered. It is shown
that for a Gaussian channel having no completely depolarizing components
the above characteristics may coincide only at non-Gaussian
mixed states and a criterion of existence of such states is given.

All the obtained results
may be reformulated as conditions for equality
between the constrained Holevo capacity of a quantum channel and the
input von Neumann entropy.
\end{abstract}
\maketitle


\section{Introduction}

The constrained Holevo capacity $\bar{C}(\Phi,\rho)$ (also called
$\chi$-function) and the quantum mutual information $I(\Phi,\rho)$
are important characteristics  of a quantum channel $\Phi$ related
respectively to the classical capacity and to the entanglement-assisted
classical capacity of this channel \cite{A&C,H-SCI,N&Ch}. These nonnegative
characteristics have the following upper bounds
\begin{equation}\label{upper-b}
    \bar{C}(\Phi,\rho)\leq H(\rho),\qquad I(\Phi,\rho)\leq 2H(\rho),
\end{equation}
where $H(\rho)$ is the von Neumann entropy of a state $\rho$, and
are connected  by the inequality
\begin{equation}\label{b-ineq}
    \bar{C}(\Phi,\rho)\leq I(\Phi,\rho).
\end{equation}

The sense of equality in the second inequality in (\ref{upper-b}) is
well known: it is equivalent to perfect reversibility of the channel
$\Phi$ on the support of the state $\rho$ \cite{H-SCI,N&Ch}. In this
paper we analyse conditions of equalities in the first inequality in
(\ref{upper-b}) and in (\ref{b-ineq}) strengthening the results
obtained in \cite{EAC}.

In fact, these inequalities are connected via the
complementary channel $\widehat{\Phi}$ to the channel $\Phi$ \cite{H-SCI,H-c-c}. This
follows from the identity
\begin{equation*}
I(\Phi,\rho)-\bar{C}(\Phi,\rho)=H(\rho)-
\bar{C}(\widehat{\Phi},\rho)
\end{equation*}
valid for any state $\rho$ with finite von Neumann entropy $H(\rho)$
(see Section 2). This identity shows, in particular, that
\begin{equation*}
\{\,\bar{C}(\Phi,\rho)=I(\Phi,\rho)<+\infty\,\}\Leftrightarrow\{\,\bar{C}(\widehat{\Phi},\rho)=H(\rho)<+\infty\,\}.
\end{equation*}
So, we may analyse conditions of equality in the first inequality in
(\ref{upper-b}) by studying conditions of equality in
(\ref{b-ineq}) and vice versa. \smallskip

This idea was used in \cite{EAC} to show that equality in
(\ref{b-ineq}) implies that the restriction of the channel $\Phi$ to
the set of all states $\sigma$ such that
$\,\supp\sigma\subseteq\supp\rho\,$ is a (discrete) c-q channel
(this was a basis step in proving the conditions for coincidence of
the Holevo capacity and the entanglement-assisted classical capacity
of a finite-dimensional channel).

In this paper we prove a simple criterion of equality in (\ref{b-ineq}) for an  infinite dimensional channel $\Phi$ and a state $\rho$ with finite (von Neumann) entropy. This criterion shows that the set of all mixed states with finite entropy, for which equality in (\ref{b-ineq}) holds, is expressed via the set $\ker\Phi$ (Theorem \ref{b-p}). It also makes it possible to prove that this equality holds for all states $\rho$ if and only if $\Phi$ is a
completely depolarizing channel (the fact conjectured in \cite{EAC}).\smallskip

We consider application of the obtained results to Bosonic Gaussian channels. In particular, we show that
for an arbitrary non-completely-depolarizing Gaussian channel $\Phi$ a strict
inequality holds in (\ref{b-ineq}) for all non-degenerate
states with finite entropy, while its validity for all mixed states with finite entropy is
equivalent to the "full rank" property of the operator describing
transformation of canonical observables.

The "complementary" results concerning an equality (strict inequality) in the first
inequality in (\ref{upper-b}) are presented in the last part of the paper.

\section{Pleriminaries}

Let $\H$ be a separable Hilbert space, $\B(\H)$ and
$\mathfrak{T}(\mathcal{H})$ -- the Banach spaces of all bounded
operators in $\mathcal{H}$ and of all trace-class operators in $\H$
correspondingly, $\S(\H)$ -- the closed convex subset of
$\mathfrak{T}(\H)$ consisting of positive operators with unit trace
called \emph{states} \cite{H-SCI,N&Ch}.

Denote by $I_{\mathcal{H}}$ and $\mathrm{Id}_{\mathcal{H}}$ the unit
operator in a Hilbert space $\mathcal{H}$ and the identity
transformation of the Banach space $\mathfrak{T}(\mathcal{H})$
correspondingly.\smallskip

Let $H(\rho)$ and $H(\rho\|\sigma)$ be respectively the von Neumann
entropy of the state $\rho$ and the quantum relative entropy of the
states $\rho$ and $\sigma$ \cite{H-SCI,N&Ch}.\smallskip

A finite or countable collection of states $\{\rho_i\}$ with the
corresponding probability distribution $\{\pi_i\}$ is called
\emph{ensemble} and denoted $\{\pi_i,\rho_i\}$. The state
$\bar{\rho}=\sum_i \pi_i\rho_i$ is called the \emph{average state}
of the ensemble $\{\pi_i,\rho_i\}$.\smallskip

The $\chi$-quantity of an ensemble $\{\pi_i,\rho_i\}$ (providing an upper bound for accessible classical information which can be obtained by applying a quantum measurement) is defined as
follows
\begin{equation}\label{chi-q}
\chi(\{\pi_i,\rho_i\})\doteq\sum_i\pi_i
H(\rho_i\|\bar{\rho})=H(\bar{\rho})-\sum_i\pi_i H(\rho_i),
\end{equation}
where the second formula is valid under the condition $H(\bar{\rho})<+\infty$ \cite{H-SCI,N&Ch}.
\smallskip

A linear completely positive trace preserving map
$\Phi:\mathfrak{T}(\mathcal{H}_A)\rightarrow\mathfrak{T}(\mathcal{H}_B)$
is called  \emph{quantum channel} \cite{H-SCI,N&Ch}. The Stinespring
theorem implies existence of a Hilbert space $\mathcal{H}_E$ and of
an isometry
$V:\mathcal{H}_A\rightarrow\mathcal{H}_B\otimes\mathcal{H}_E$ such
that
\begin{equation}\label{Stinespring-rep}
\Phi(\rho)=\mathrm{Tr}_{\mathcal{H}_E}V\rho V^{*},\quad
\rho\in\mathfrak{T}(\mathcal{H}_A).
\end{equation}
The quantum  channel
\begin{equation}\label{c-channel}
\mathfrak{T}(\mathcal{H}_A)\ni
\rho\mapsto\widehat{\Phi}(\rho)=\mathrm{Tr}_{\mathcal{H}_B}V\rho V^{*}\in\mathfrak{T}(\mathcal{H}_E)
\end{equation}
is called \emph{complementary} to the channel $\Phi$
\cite{H-SCI,H-c-c}. The complementary channel is defined uniquely: if
$\widehat{\Phi}':\mathfrak{T}(\mathcal{H}_A)\rightarrow\mathfrak{T}(\mathcal{H}_{E'})$
is a channel defined by (\ref{c-channel}) via the Stinespring
isometry
$V':\mathcal{H}_A\rightarrow\mathcal{H}_B\otimes\mathcal{H}_{E'}$
then the channels $\widehat{\Phi}$ and $\widehat{\Phi}'$ are isometrically equivalent in the sense of the
following definition \cite{H-c-c}.\smallskip

\begin{definition}\label{isom-eq}
Channels
$\Phi:\mathfrak{T}(\mathcal{H}_A)\rightarrow\mathfrak{T}(\mathcal{H}_B)$
and
$\,\Phi':\mathfrak{T}(\mathcal{H}_{A})\rightarrow\mathfrak{T}(\mathcal{H}_B')\,$
are \emph{isometrically equivalent} if there exists a partial
isometry $W:\mathcal{H}_B\rightarrow\mathcal{H}_{B'}$ such that
\begin{equation*}
\Phi'(\rho)=W\Phi(\rho)W^*,\quad\Phi(\rho)=W^*\Phi'(\rho)W,\quad
\rho\in \T(\H_A).
\end{equation*}
\end{definition}

Throughout the paper we will use the following notion.
\smallskip

\begin{definition}\label{subch}
The restriction of a channel $\Phi:\T(\H_A)\rightarrow\T(\H_B)$ to the
set $\T(\H_0)$, where $\H_0$ is a nontrivial subspace of $\H_A$, is
called \emph{subchannel of $\,\Phi$ corresponding to the subspace
$\H_0$}.
\end{definition} \smallskip

By definition the complementary channel to the subchannel of an
arbitrary  channel $\Phi$ corresponding to any subspace $\H_0$
coincides with the subchannel of the complementary channel
$\widehat{\Phi}$ corresponding to the subspace $\H_0$, i.e.
$\widehat{\Psi}=\widehat{\Phi}|_{\T(\H_0)}$, where
$\Psi=\Phi|_{\T(\H_0)}$. \medskip

The following class of quantum channels  plays a basic role in this
paper.\smallskip

\begin{definition}\label{d-c-q}
A channel $\Phi:\T(\H_A)\rightarrow\T(\H_B)$ is called
\emph{classical-quantum of discrete type} (briefly, \emph{discrete
c-q channel}) if it has the following representation
\begin{equation}\label{c-q-rep}
\Phi(\rho)=\sum_{i=1}^{\dim\H_A}\langle
i|\rho|i\rangle\sigma_i,\quad \rho\in \T(\H_A),
\end{equation}
where $\{|i\rangle\}$ is an orthonormal basis in $\H_A$ and
$\{\sigma_i\}$ is a collection of states in $\S(\H_B)$.\footnote{In infinite dimensions there exist c-q channels of non-discrete type \cite[the Appendix]{H-Sh-3}.}
\end{definition}\smallskip

Discrete c-q channel (\ref{c-q-rep}) such that $\sigma_i=\sigma$
for all $\,i\,$ is a \emph{completely depolarizing} channel
$\Phi(\rho)=[\Tr\rho]\sigma$. We will use the following obvious
lemma.\smallskip
\begin{lemma}\label{cdc}
\emph{A channel $\,\Phi:\T(\H_A)\rightarrow\T(\H_B)$ is completely depolarizing if and only if
$\,\Phi(|\varphi\rangle\langle\psi|)=0$ for any orthogonal vectors
$\varphi,\psi\in\H_A$.}
\end{lemma}\smallskip

For given channel $\Phi$ and ensemble $\{\pi_i,\rho_i\}$ the
$\chi$-quantity of the ensemble $\{\pi_i,\Phi(\rho_i)\}$ will be
denoted $\chi_{\Phi}(\{\pi_i,\rho_i\})$.

The constrained Holevo capacity of a channel $\Phi:\T(\H_A)\rightarrow\T(\H_B)$ at a state
$\rho\in\S(\H_A)$ is defined as follows
\begin{equation}\label{chi-fun}
\bar{C}(\Phi,\rho)=\sup_{\sum_i\pi_i\rho_i=\rho}\chi_{\Phi}(\{\pi_i,\rho_i\}),
\end{equation}
where the supremum is over all finite or countable ensembles
$\{\pi_i,\rho_i\}$ with the average state $\rho$ \cite{H-SCI,H-Sh-2}.\footnote{In \cite{H-Sh-2,EAC} the
constrained Holevo capacity $\bar{C}(\Phi,\rho)$ is denoted
$\chi_{\Phi}(\rho)$ and called the
$\chi$\nobreakdash-\hspace{0pt}function of the channel $\Phi$.}
If $H(\Phi(\rho))<+\infty$ then
\begin{equation}\label{chi-fun+}
\bar{C}(\Phi,\rho)=H(\Phi(\rho))-\hat{H}_{\Phi}(\rho),
\end{equation}
where $\hat{H}_{\Phi}(\rho)=\inf_{\sum_i\pi_i\rho_i=\rho}\sum_i\pi_i
H(\Phi(\rho_i))$ is the $\sigma$-convex hull of the function
$\rho\mapsto H(\Phi(\rho))$. Note that the above supremum and
infimum can be taken over ensembles of pure states.\smallskip

Since monotonicity of the quantum relative entropy implies
$$
\chi_{\Phi}(\{\pi_i,\rho_i\})\leq\chi(\{\pi_i,\rho_i\})\leq H(\rho)
$$
for any ensemble $\{\pi_i,\rho_i\}$ with the average state $\rho$,
the inequality
\begin{equation}\label{b-ineq-2}
    \bar{C}(\Phi,\rho)\leq H(\rho)
\end{equation}
holds for an arbitrary quantum channel $\Phi$ and any state $\rho$.
\medskip

The quantum mutual information of a finite-dimensional channel
$\Phi$ at a state $\rho\in\mathfrak{S}(\H_A)$ is defined by
one of the formulas
\begin{equation}\label{mi}
I(\Phi,\rho)=H(\rho)+H(\Phi(\rho))-H(\Phi,\rho)=H(\rho)+H(\Phi(\rho))-H(\widehat{\Phi}(\rho)),
\end{equation}
where $H(\Phi,\rho)$ is the entropy exchange of the channel
$\Phi$ at the state $\rho$ \cite{A&C,H-SCI}.

In infinite dimensions this definition may contain uncertainty
$"\infty-\infty"$, but it can be modified to avoid this problem as
follows
\begin{equation*}
I(\Phi,\rho) = H\left(\Phi \otimes \id_{\H_R}
(\hspace{1pt}\hat{\rho}\hspace{2pt})\hspace{1pt} \|\hspace{1pt} \Phi \otimes \id_{\H_R} (\hspace{1pt}\rho\otimes
\varrho\,)\right),
\end{equation*}
where $\H_R$ is a Hilbert space isomorphic to $\H_A$,
$\,\hat{\rho}\,$ is a purification\footnote{This means that
$\Tr_{\H_R}\hat{\rho}=\rho$.} of the state $\rho$ in the space
$\H_A\otimes\H_R$ and $\,\varrho=\Tr_{\H_A}\hat{\rho}\,$ is a state in
$\S(\H_R)$ isomorphic to $\rho$ \cite{H-SCI}. \medskip

A basic role in this paper is played by the expression
\begin{equation}\label{b-rep+}
I(\Phi,\rho)=H(\rho)+ \bar{C}(\Phi,\rho)-
\bar{C}(\widehat{\Phi},\rho)
\end{equation}
valid for an arbitrary quantum channel $\Phi$ and any state $\rho$
with finite entropy (the condition $H(\rho)<+\infty$ implies
finiteness of all the terms in (\ref{b-rep+})). If
$H(\Phi(\rho))<+\infty$ and (hence)
$H(\widehat{\Phi}(\rho))<+\infty$ then this expression is easily
proved by using (\ref{chi-fun+}) and by noting that
$\hat{H}_{\Phi}\equiv\hat{H}_{\widehat{\Phi}}$ (this follows from
coincidence of the functions $\rho\mapsto H(\Phi(\rho))$ and
$\rho\mapsto H(\widehat{\Phi}(\rho))$ on the set of pure states). In
general case expression (\ref{b-rep+}) follows from Proposition 4 in \cite{H-Sh-3}.\smallskip

The constrained Holevo capacity and the quantum mutual information
of an arbitrary channel $\Phi$ at any state $\rho$ are connected by
the inequality
\begin{equation}\label{b-ineq-1}
    \bar{C}(\Phi,\rho)\leq I(\Phi,\rho).
\end{equation}
If $H(\rho)<+\infty$ then this inequality directly follows from
(\ref{b-ineq-2}) and (\ref{b-rep+}). For an arbitrary state $\rho$
it can be proved by using  the sequence  of finite rank
states $\rho_n=[\Tr
P_n\rho]^{-1}P_n\rho$, where $P_n$ is the spectral projector of $\rho$ corresponding to its $n$ maximal eigenvalues.  Concavity and lower
semicontinuity of the functions $\rho\mapsto\bar{C}(\Phi,\rho)$ and
$\rho\mapsto I(\Phi,\rho)$ imply respectively
\begin{equation}\label{lim-rel}
\lim_n\bar{C}(\Phi,\rho_n)=\bar{C}(\Phi,\rho)\leq+\infty,\quad\lim_nI(\Phi,\rho_n)=I(\Phi,\rho)\leq+\infty.
\end{equation}
Thus, validity of (\ref{b-ineq-1}) for the state $\rho$ follows from
validity of (\ref{b-ineq-1}) for each state of the sequence
$\{\rho_n\}$.\smallskip

Expression (\ref{b-rep+}) shows that under the condition
$H(\rho)<+\infty$ inequalities (\ref{b-ineq-2}) and (\ref{b-ineq-1})
are, roughly speaking, complementary to each other, in particular,
\begin{equation}\label{b-e-r+}
\{\,\bar{C}(\Phi,\rho)=I(\Phi,\rho)<+\infty\,\}\;\Leftrightarrow\;\{\,\bar{C}(\widehat{\Phi},\rho)=H(\rho)<+\infty\,\}.
\end{equation}
and hence we may analyse conditions of equality in (\ref{b-ineq-1})
by studying conditions of equality in (\ref{b-ineq-2}).\smallskip

Relation (\ref{b-e-r+}) implies the following result essentially used below.\smallskip
\begin{lemma}\label{chain} \emph{Let $\,\Phi:\T(\H_A)\rightarrow\T(\H_B)$ and $\,\Psi:\T(\H_B)\rightarrow\T(\H_C)$ be quantum
channels and $\rho$ a state in $\S(\H_A)$ such that
$H(\rho)<+\infty$. Then}
$$
\bar{C}(\Phi,\rho)=I(\Phi,\rho)\quad\Rightarrow\quad\bar{C}(\Psi\circ\Phi,\rho)=I(\Psi\circ\Phi,\rho).
$$
\end{lemma}

\textbf{Proof.} It follows from the proof of Lemma 17 in \cite{R} that
there exists a channel $\Theta$ such that
$\widehat{\Phi}=\Theta\circ\widehat{\Psi\circ\Phi}$.\footnote{This fact can be also shown easily by using the representation of a complementary channel via the Kraus operators of initial channel \cite[formula (11)]{H-c-c}.} Hence the chain
rule for the Holevo capacity and (\ref{b-ineq-2}) show that
$$
\bar{C}(\widehat{\Phi},\rho)=H(\rho)\quad\Rightarrow\quad\bar{C}(\widehat{\Psi\circ\Phi},\rho)=H(\rho).
$$
By (\ref{b-e-r+}) this implication is equivalent to the assertion of
the lemma. $\square$\smallskip

We will study the case of equality in (\ref{b-ineq-2}) by using the
concept of reversibility (sufficiency) of a quantum channel with
respect to families of input states \cite{J&P,J-rev,P-sqc}. A
channel $\Phi:\T(\H_A)\rightarrow\T(\H_B)$ is called reversible
(sufficient) with respect to a family $\S\subseteq\S(\H_A)$ if there
is a channel $\,\Psi:\T(\H_B)\rightarrow\T(\H_A)$ such that
$\,\rho=\Psi\circ\Phi(\rho)\,$ for all $\,\rho\in\S$. The
characterization of reversibility is obtained in \cite{J&P} (by
generalizing the results of \cite{P-sqc}). We will use the following
implication of this characterization.\smallskip

\begin{theorem}\label{petz} \emph{Let $\,\Phi:\T(\H_A)\rightarrow\T(\H_B)$ be a quantum
channel and $\,\S=\{\rho_i\}$ a family of states in $\,\S(\H_A)$. Let
$\,\{\pi_i\}$ be a non-degenerate probability distribution such that
$\,\chi(\{\pi_i,\rho_i\})<+\infty$. The channel $\,\Phi$ is reversible
with respect to the family $\,\S$ if and only if}
$$
\chi_{\Phi}(\{\pi_i,\rho_i\})=\chi(\{\pi_i,\rho_i\}).
$$
\end{theorem}\smallskip

For our purposes we will need a special case in which the family $\S$
consists of pure states. Necessary and sufficient conditions for
reversibility of a quantum channel with respect to families
of pure states are considered in Section 3 in \cite{BRC}. We will use
the following implications of that results.\smallskip

\begin{theorem}\label{rc} \emph{Let $\,\Phi:\T(\H_A)\rightarrow\T(\H_B)$ be a quantum
channel and $\,\S=\{|\varphi_i\rangle\langle\varphi_i|\}$ a family
of pure states in $\,\S(\H_A)$.}\smallskip

A) \emph{If the family $\,\S$ consists of orthogonal states then the
channel $\,\Phi$ is reversible with respect to the family $\,\S$ if
and only if
\begin{equation*}
\widehat{\Phi}(\rho)=\sum_{i=1}^{\dim\H_{\S}}\langle
\varphi_i|\rho|\varphi_i\rangle\sigma_i\quad \forall\rho\in
\S(\H_{\S}),
\end{equation*}
where $\H_{\S}$ is the subspace of $\,\H_A$ generated by the
family $\,\{|\varphi_i\rangle\}$ and $\{\sigma_i\}$ is a collection of states in
$\,\S(\H_B)$.}\smallskip

B) \emph{Let $\S=\bigcup_k\S_k$ be a  decomposition of $\,\S$ into
disjoint orthogonally non-decomposable subfamilies \footnote{This
means  that $\rho\perp\sigma$ if $\rho\in \S_k$ and $\sigma\in \S_l$
for all $k\neq l$ and that for each $k$ there is no subspace $\H_0$
such that some states (not all!) from $\S_k$ lie in $\H_0$, while
the others -- in $\H^{\perp}_0$ \cite{BRC}.} and $\,P_k$ the
projector on the subspace generated by all the states in $\,\S_k$.
If the channel $\,\Phi$ is reversible with respect to the family
$\S$ then it is reversible with respect to the family}
  $$
  \widehat{\S}=\left\{\rho\in\S(\H_A)\,\left|\;\rho=\sum_k P_k\rho P_k\right.\right\}.
  $$\vspace{-15pt}
\smallskip
\end{theorem}

Theorem \ref{rc} shows, in particular, that reversibility of a
quantum channel with respect to at least one family of pure states
is equivalent to existence of at least one discrete c-q subchannel
of the complementary channel. A simple criterion of the last
property is given by the following lemma (which can be proved
similarly to Lemma 3 in \cite{BRC}).\smallskip

\begin{lemma}\label{c-q} \emph{A channel $\,\Phi:\T(\H_A)\rightarrow\T(\H_B)$ has a discrete c-q subchannel if and
only if there exists an orthogonal family $\{|i\rangle\}$ of
unit vectors in $\H_A$ such that $\,\Phi(|i\rangle\langle j|)=0$ for all
$\,i\neq j$. In this case the subchannel $\,\Phi|_{\T(\H_0)}$ has
representation (\ref{c-q-rep}) with $\H_0=\overline{\mathrm{lin}}\{|i\rangle\}$
instead of $\,\H_A$.}

\end{lemma}

\section{The main results}

By using Theorem \ref{petz} and equivalence relation (\ref{b-e-r+}) it is shown in \cite{EAC} that 
\begin{equation*}
    \bar{C}(\Phi,\rho)=I(\Phi,\rho)\quad\Rightarrow\quad\Phi|_{\T(\H_{\rho})}\;\,\textup{is a discrete c-q
    channel},
\end{equation*}
for a finite-dimensional channel
$\Phi$, where $\H_{\rho}$ is the support of the state $\rho$. Theorem
\ref{rc} makes it possible to strengthen this observation by showing
that $\Phi|_{\T(\H_{\rho})}$ is a discrete c-q  channel
\emph{determined by a particular basis of eigenvectors of the state $\rho$}
\footnote{Here and in what follows speaking about basis of
eigenvectors of a state we mean a basis in the support of this
state.} and to generalize it to infinite dimensions. As a result we
obtain the following criterion of an equality in (\ref{b-ineq-1})
for an infinite-dimensional channel $\Phi$ and a state
$\rho$ with finite von Neumann entropy.\smallskip

\begin{theorem}\label{b-p}
\emph{Let $\,\Phi:\T(\H_A)\rightarrow\T(\H_B)$ be a quantum channel
and $\,\Pi(\Phi)$ the set of all orthogonal families
$\{|i\rangle\}$ of unit vectors in $\H_A$ such that $\,\Phi(|i\rangle\langle j|)=0$
for all $\,i\neq j$.}
\medskip

\noindent A) \emph{Let $\rho$ be a mixed state in $\S(\H_A)$ such that
$H(\rho)<+\infty$. The following statements are equivalent:}
\begin{enumerate}[(i)]
    \item \emph{$\bar{C}(\Phi,\rho)=I(\Phi,\rho)$;}
    \item \emph{the set $\,\Pi(\Phi)$ contains at least one basis of
eigenvectors of the state $\rho$;}
    \item \emph{$\Phi(\varrho)=\sum_i\langle\varphi_i|\varrho|\varphi_i\rangle\sigma_i$
for any $\varrho\in \S(\H_{\rho})$, where $\H_{\rho}=\supp\rho$,
$\{|\varphi_i\rangle\}$ is a particular orthonormal basis of
eigenvectors of the state $\rho$ and $\{\sigma_i\}$ is a collection
of states in $\S(\H_B)$.}
\end{enumerate}
\emph{For a state $\rho$ with infinite entropy the above statements
are related as follows
$\mathrm{(ii)\Leftrightarrow(iii)\Rightarrow(i)}$ (with possible
infinite values of the both sides of the equality in
$\mathrm{(i)}$)}.\medskip

\noindent B) \emph{The set $\,\S_{\Phi}^{=}$  of all mixed states
$\rho$ in $\S(\H_A)$ with finite entropy, for which $\mathrm{(i)}$
holds, can be represented as follows
\begin{equation}\label{s-rep}
\S_{\Phi}^{=}\,=\!\bigcup_{\{|i\rangle\}\in\Pi(\Phi)}\left\{\left.\rho=\sum_i\pi_i|i\rangle\langle
i|\,\,\right|\,\{\pi_i\}\in\mathfrak{P}_{\mathrm{f}}\;\right\},\!
\end{equation}
where $\mathfrak{P}_{\mathrm{f}}$ is the set of probability
distributions with finite Shannon entropy.}

\end{theorem}\medskip

\begin{remark}\label{b-p-r-2}
By Theorem \ref{b-p}B the set $\S_{\Phi}^{=}$ is completely
determined by the set $\ker\Phi$ (since the set $\Pi(\Phi)$ is
defined via the set $\ker\Phi$). The example of a channel $\Phi$,
for which the set $\Pi(\Phi)$ contains infinitely many different
non-complete families of vectors, is considered in the next section
(where a description of the set $\Pi(\Phi)$ for Bosonic Gaussian
channels is given).
\end{remark}\medskip

\begin{remark}\label{b-p-r-2}
The condition $H(\rho)<+\infty$ is essentially used in the proof of
the implication $\mathrm{(i)\Rightarrow(ii),(iii)}$ (since it is
based on relation (\ref{b-e-r+})) and it is not clear how to relax
it. On the other hand, this condition seems technical and one can
conjecture that the assertions of Theorem \ref{b-p} and of Corollary
\ref{b-p-c-1} below are also valid for states with infinite
entropy.\footnote{I would be grateful for any comments concerning
this question.}
\end{remark}\medskip

\textbf{Proof.} A) Note first that $\mathrm{(ii)\Leftrightarrow(iii)}$ follows from Lemma \ref{c-q}.
\smallskip

$\mathrm{(iii)\Rightarrow(i)}$. By Theorem \ref{rc}A
$\mathrm{(iii)}$ implies reversibility of the channel $\widehat{\Phi}$  with respect to the family
$\{|\varphi_i\rangle\langle\varphi_i|\}$ which shows that $\bar{C}(\widehat{\Phi},\rho)=H(\rho)$.
So, if $H(\rho)<+\infty$ then $\mathrm{(i)}$ follows from (\ref{b-e-r+}).

If $H(\rho)=+\infty$ then this reversibility implies
$\bar{C}(\widehat{\Phi},\rho_n)=H(\rho_n)$ and hence
$\bar{C}(\Phi,\rho_n)=I(\Phi,\rho_n)$, where $\rho_n=[\Tr
P_n\rho]^{-1}P_n\rho$,
$P_n=\sum_{i=1}^n|\varphi_i\rangle\langle\varphi_i|$. Relations
(\ref{lim-rel}) show that
$\bar{C}(\Phi,\rho)=I(\Phi,\rho)\leq+\infty$.
\smallskip

$\mathrm{(i)\Rightarrow(iii)}$ Here we prove this implication
assuming that $\Phi$ is a finite-dimensional channel
($\dim\H_A,\dim\H_B<+\infty$). A general proof is based on the same
idea but requires technical results (related to the notion
of a generalized ensemble) and additional approximation, it
is presented in the Appendix.

If $\Phi$ is a finite-dimensional channel then $\widehat{\Phi}$ is a
finite-dimensional channel as well (cf.\cite{H-c-c}) and hence for
any state $\rho\in\S(\H_A)$ the supremum in the expression for
$\bar{C}(\widehat{\Phi},\rho)$ (expression (\ref{chi-fun}) with
$\widehat{\Phi}$ instead of $\Phi$) is achieved at a particular
finite ensemble $\{\pi_i,\rho_i\}$ of pure states, i.e.
$$
\bar{C}(\widehat{\Phi},\rho)=\chi_{\widehat{\Phi}}(\{\pi_i,\rho_i\}),\quad\sum_i\pi_i\rho_i=\rho
$$
(existence of such ensemble can be proved by using the arguments from
\cite{Sch-West}).\smallskip

If $\mathrm{(i)}$ holds then (\ref{b-e-r+}) implies
$\bar{C}(\widehat{\Phi},\rho)=H(\rho)=\chi(\{\pi_i,\rho_i\})$ and
hence
$\chi_{\widehat{\Phi}}(\{\pi_i,\rho_i\})=\chi(\{\pi_i,\rho_i\})$. By
Theorem \ref{petz} this is equivalent to reversibility of the
channel $\widehat{\Phi}$ with respect to the family
$\S=\{\rho_i\}$. Let $\S=\bigcup_k\S_k$
be a decomposition of $\S$ into disjoint
non-orthogonally-decomposable subfamilies (see Theorem \ref{rc}B).
Denote by $I_k$ the set of all indexes $i$ such that
$\rho_i\in\S_k$. Let
$\{|\varphi_k^i\rangle\}_i$ be an orthonormal basis of eigenvectors
of the positive operator $\rho_k=\sum_{i\in
I_k}\pi_i\rho_i$. Since $\rho=\sum_k\rho_k$ and 
$\,\supp\rho_k\perp\supp\rho_l\,$ for all $k\neq l$, $\{|\varphi_k^i\rangle\}_{ik}$ is an
orthonormal basis of eigenvectors of the state $\rho$.

By Theorem \ref{rc}B the reversibility of the channel
$\widehat{\Phi}$ with respect to the family $\S$ implies
reversibility of this channel with respect to the orthogonal family
$\{|\varphi_k^i\rangle\langle\varphi_k^i|\}_{ik}$ (since the last
family is contained in the family $\widehat{\S}$). Since the channels $\Phi$ and
$\widehat{\widehat{\Phi}}$ are isometrically equivalent (see
Definition \ref{isom-eq}), Theorem \ref{rc}A implies
$\mathrm{(iii)}$.\smallskip

B) Representation (\ref{s-rep}) directly follows from the first part of the theorem. $\square$\medskip

Theorem \ref{b-p} implies  sufficient
conditions for a strict inequality in (\ref{b-ineq-1}).\smallskip

\begin{corollary}\label{b-p-c-1} \emph{Let $\,\Phi:\T(\H_A)\rightarrow\T(\H_B)$ be a quantum
channel.}\smallskip

A) \emph{If $\,\Phi$ is not a discrete c-q channel  then
$\bar{C}(\Phi,\rho)<I(\Phi,\rho)$ for any non-degenerate state
$\rho$ in $\S(\H_A)$ with finite  entropy.}\smallskip

B) \emph{If the set $\,\ker\Phi$ contains no 1-rank operators then
$\bar{C}(\Phi,\rho)<I(\Phi,\rho)$ for any mixed state $\rho$ in
$\S(\H_A)$ with finite  entropy.}\smallskip
\end{corollary}\medskip

The examples of a channel for which the condition of Corollary
\ref{b-p-c-1}B holds are considered in the next section (Proposition
\ref{b-l-c} and Example \ref{e-1}).\medskip

Now we turn to a condition of global equality in (\ref{b-ineq-1}) and
prove the strengthened version of the conjecture stated in
\cite{EAC}.\smallskip

\begin{corollary}\label{b-p+}
\emph{If $\,\bar{C}(\Phi,\rho)=I(\Phi,\rho)\,$ for any 2-rank state
$\rho$ then $\,\Phi$ is a completely depolarizing channel and hence
$\,\bar{C}(\Phi,\rho)=I(\Phi,\rho)=0\,$ for any state
$\rho$.}\smallskip
\end{corollary}

\begin{remark}\label{b-p+r}
Corollary \ref{b-p+} shows that for any nontrivial channel $\Phi$
the concave nonnegative functions $\rho\mapsto\bar{C}(\Phi,\rho)$
and $\rho\mapsto I(\Phi,\rho)$ (equal to zero on the set of 1-rank
states) are always separated by a particular 2-rank
state.\smallskip
\end{remark}

\textbf{Proof.} By Lemma \ref{cdc} it suffices to show that
$\Phi(|\varphi\rangle\langle\psi|)=0$ for any orthogonal unit vectors
$\varphi,\psi\in\H_A$.

Let
$\rho=0.3|\varphi\rangle\langle\varphi|+0.7|\psi\rangle\langle\psi|$
be a 2-rank state in $\S(\H_A)$. By the condition
$\bar{C}(\Phi,\rho)=I(\Phi,\rho)$ and Theorem \ref{b-p} implies
$\Phi(|\varphi\rangle\langle\psi|)=0$ (since
$\{|\varphi\rangle,|\psi\rangle\}$ is the only basis of eigenvectors
of the state $\rho$). $\square$ \medskip

As shown in \cite{EAC} for a finite-dimensional channel
$\Phi:\T(\H_A)\rightarrow\T(\H_B)$ the following relation holds
$$
D(\Phi)\doteq\max_{\rho\in\S(\H_A)}\left[\,I(\Phi,\rho)-\bar{C}(\Phi,\rho)\,\right]=
\sup_{H,h}
\left[\,C_{\mathrm{ea}}(\Phi,H,h)-\bar{C}(\Phi,H,h)\,\right],
$$
where $C_{\mathrm{ea}}(\Phi,H,h)$ and $\bar{C}(\Phi,H,h)$ are
respectively the entanglement-assisted capacity and the Holevo
capacity of the channel $\Phi$ with the linear constraint determined
by inequality $\Tr H\rho\leq h$ and the supremum is taken over all
pairs (positive operator $H$, positive number $h$). Corollary \ref{b-p+}
implies that $D(\Phi)>0$ if $\Phi$ is not completely
depolarizing. This completes the proof of the following list of
properties of the parameter $D(\Phi)$ (showing that it can be
considered as one of characteristics of the channel $\Phi$
describing its "level of noise"):
\begin{itemize}
    \item $D(\Psi\circ\Phi)\leq D(\Phi)$ for any channel $\Psi:\T(\H_B)\rightarrow\T(\H_C)$;
    \item $D(\Phi)\in[0,\log\dim\H_A]$;
    \item $D(\Phi)=\log\dim\H_A$ if and only if $\Phi$ is a noiseless channel (i.e. $\Phi$ is unitary equivalent to the channel $\rho\mapsto\rho\otimes\sigma$,
where $\sigma$ is a given state);
    \item $D(\Phi)=0\,$ if and only if $\Phi$ is a completely depolarizing channel.
\end{itemize}

\section{Applications to Bosonic Gaussian channels}

Consider application of Theorem \ref{b-p} to Bosonic Gaussian channels playing
a central role in infinite-dimensional quantum information theory \cite{E&W,H-SCI}.

Let $\mathcal{H}_{X}$ $(X=A,B,...)$ be the space of irreducible
representation of the Canonical Commutation Relations (CCR)
\begin{equation*}
W_X(z)W_X(z^{\prime })=\exp \left(-\frac{i}{2}z^{\top }\Delta
_{X}z^{\prime }\right) W_X(z^{\prime }+z)  
\end{equation*}
with a symplectic space $(Z_{X},\Delta _{X})$ and the Weyl operators
$W_{X}(z)$ \cite[Ch.12]{H-SCI}. Denote by $s_X$ the number of modes
of the system $X$, i.e. $2s_X=\dim Z_X$.\smallskip

A Bosonic Gaussian channel  $\Phi
:\mathfrak{T}(\mathcal{H}_{A})\rightarrow
\mathfrak{T}(\mathcal{H}_{B})$ is defined via the action of its dual
$\Phi^{\ast }:\mathfrak{B}(\mathcal{H}_{B})\rightarrow
\mathfrak{B}(\mathcal{H}_{A})$ on the Weyl operators:
\begin{equation}\label{g-ch-def}
\Phi^{\ast}(W_{B}(z))=W_A(Kz)\exp \left[ il^{\top
}z-\textstyle\frac{1}{2} z^{\top }\alpha z\right],\quad z\in Z_B,
\end{equation}
where $K$ is a linear operator $Z_{B}\rightarrow Z_{A}$, $\,l\,$ is a
$\,2s_B$-dimensional real row and $\,\alpha\,$ is a real symmetric
$\,(2s_B)\times(2s_B)$ matrix satisfying the inequality
\begin{equation}\label{nid}
\alpha \geq \pm \frac{i}{2}\left[ \Delta _{B}-K^{\top }\Delta
_{A}K\right].
\end{equation}

By applying replacement unitary transformations an arbitrary Gaussian
channel can be transformed to the Gaussian channel with $l=0$ and
the same matrices $K$ and $\alpha$ (such channel is called
\emph{centered} and will be denoted $\Phi_{K,\alpha}$). So, in study of relations between the
constrained Holevo capacity and the quantum mutual information we
may (and will) assume that \textbf{all Gaussian channels are
centered}.\smallskip

It follows from Proposition 5 in \cite{H-Sh-3} and Proposition 3 in \cite{BRC} that:
\begin{itemize}
    \item $\Phi_{K,\alpha}$ is a discrete c-q channel if and only
    if $K=0$ (i.e. $\Phi_{K,\alpha}$ is completely depolarizing);
    \item $\Phi_{K,\alpha}$ has discrete c-q subchannels if and only if $\mathrm{Ran}K\neq Z_A$ (i.e.
    $\rank K<\dim Z_A$).
\end{itemize}

Let $\Phi_{K,\alpha}$ be a nontrivial Gaussian channel $(K\neq0)$.
By the above observations Theorem \ref{b-p} shows that the strict
inequality $\bar{C}(\Phi_{K,\alpha},\rho)<I(\Phi_{K,\alpha},\rho)$
is valid for any non-degenerate state $\rho$ with finite entropy while
existence of mixed degenerate states, for which
an equality holds in this inequality, is possible only if
$\mathrm{Ran}K\neq Z_A$. This condition holds in the following two
cases:
\begin{enumerate}[A)]
    \item $[\mathrm{Ran} K]^{\perp}$ is a nontrivial isotropic subspace of $Z_A$; \footnote{$[\mathrm{Ran} K]^{\perp}$ is the skew-orthogonal
complementary subspace to the subspace $\mathrm{Ran} K$ of $Z_A$. We will
use this sense of the symbol $"\perp"$ dealing with a
subspace of a symplectic space.}
    \item $[\mathrm{Ran} K]^{\perp}$ contains a nontrivial symplectic
    subspace.
\end{enumerate}

By Proposition 3 in \cite{BRC} Gaussian channels corresponding to
case B are characterized by existence of completely depolarizing
subchannels. The proof of Proposition \ref{b-l-c} below shows that
any such channel can be represented as a partial trace over some
input modes followed by a Gaussian channel which either corresponds
to case A or satisfies the condition $"\mathrm{Ran}K=Z_A"$. So, we
will focus attention on case A and will find all mixed states $\rho$
with finite entropy such that
$\bar{C}(\Phi_{K,\alpha},\rho)=I(\Phi_{K,\alpha},\rho)$ by
describing the set $\Pi(\Phi_{K,\alpha})$ (introduced in Theorem
\ref{b-p}) in the Schrodinger representation.\smallskip

Since the family $\{W_B(z)\}_{z\in Z_B}$ generates $\B(\H_B)$, it
follows from definition (\ref{g-ch-def}) of the channel
$\Phi_{K,\alpha}$ that $\Phi_{K,\alpha}(|\varphi\rangle\langle
\psi|)=0$ for vectors $\varphi,\psi\in\H_A$ if and only if $\langle\varphi|W_A(Kz)|\psi\rangle=0$ for all
$z\in Z_B$. So, the set $\Pi(\Phi_{K,\alpha})$ consists of all orthonormal families $\{|\varphi_i\rangle\}\subset\H_A$ such that  
\begin{equation}\label{s-eq-rel}
\langle\varphi_i|W_A(Kz)|\varphi_j\rangle=0\;\;\;\forall
z\in Z_B,\;\forall i\neq j.
\end{equation}
In case A Lemma 6 in \cite[Appendix 6.2]{BRC} implies existence of a
symplectic basis $\{\tilde{e}_k, \tilde{h}_k\}$ in $Z_A$ such that
$\{\tilde{e}_1,...,\tilde{e}_{s_A},
\tilde{h}_{d+1},...,\tilde{h}_{s_A}\}$ is a basis in
$\mathrm{Ran}K$, $d\leq s_A$. Let $Z_B^0$ be a subspace of $Z_B$
with the basis $\{z^e_1,...,z^e_{s_A},z^h_{d+1},...,z^h_{s_A} \}$
such that $\tilde{e}_k=Kz^e_k$ for all $k=\overline{1,{s_A}}$ and
$\tilde{h}_k=Kz^h_k$ for all $k=\overline{d+1,{s_A}}$. Thus for any
vector $z\in Z_B^0$ represented as $z=\sum_{k=1}^{s_A} x_k
z^e_k+\sum_{k=d+1}^{s_A} y_k z^h_k$,
$(x_1,...,x_{s_A})\in\mathbb{R}^{s_A}$,
$(y_{d+1},...,y_{s_A})\in\mathbb{R}^{s_A-d}$ we have
$$
\begin{array}{c}
   \displaystyle W_A(Kz)=W_A\left(\sum_{k=1}^{s_A} x_k
\tilde{e}_k+\sum_{k=d+1}^{s_A} y_k \tilde{h}_k\right)\medskip\\
   \displaystyle =\lambda W_A(x_1
\tilde{e}_1)\cdot...\cdot W_A(x_{s_A} \tilde{e}_{s_A})\cdot
W_A(y_{d+1} \tilde{h}_{d+1})\cdot...\cdot W_A(y_{s_A}
\tilde{h}_{s_A}),
 \end{array}
$$
where $\lambda=e^{i [x_{d+1}y_{d+1}+...+x_{s_A}y_{s_A}}]\neq 0$.

By identifying the space $\H_A$ with the space
$L_2(\mathbb{R}^{s_A})$ of complex-valued functions of $s_A$
variables (which will be denoted $\xi_1,...,\xi_{s_A}$) and the Weyl
operators $W_A(\tilde{e}_k)$ and $W_A(\tilde{h}_k)$ with the
operators
$$
\psi(\xi_1,...,\xi_{s_A})\mapsto e^{i\xi_k}
\psi(\xi_1,...,\xi_{s_A})\;\;\,\text{and}\;\;
\psi(\xi_1,...,\xi_{s_A})\mapsto
\psi(\xi_1,...,\xi_{k}+1,...,\xi_{s_A})
$$
the equality in (\ref{s-eq-rel}) for the
vector $z$ can be written as follows
\begin{equation*}
\begin{array}{c}
\displaystyle\int \overline{\varphi_i(\xi_1,...,\xi_{s_A})}
(S_{y_{d+1},...,y_{s_A}}\varphi_j)(\xi_1,...,\xi_{s_A}) e^{i (x_1\xi_1+...+x_{s_A}\xi_{s_A})}d\xi_1,...,d\xi_{s_A}=0,\\\\
\end{array}
\end{equation*}
where
$(S_{y_{d+1},...,y_{s_A}}\varphi_j)(\xi_1,...,\xi_{s_A})=\varphi_j(\xi_1,...,\xi_d,\xi_{d+1}+y_{d+1},...,\xi_{s_A}+y_{s_A})$.\medskip

This equality holds for all $(x_1,...,x_{s_A})\in\mathbb{R}^{s_A}$
and $(y_{d+1},...,y_{s_A})\in\mathbb{R}^{s_A-d}$ (that is for all
$z\in Z_B^0$) if and only if
$$
\overline{\varphi_i(\xi_1,...,\xi_{s_A})}
(S_{y_{d+1},...,y_{s_A}}\varphi_j)(\xi_1,...,\xi_{s_A})=0
$$
for almost all $(\xi_1,...,\xi_{s_A})\in\mathbb{R}^{s_A}$ and all
$(y_{d+1},...,y_{s_A})\in\mathbb{R}^{s_A-d}$. Since
$\mathrm{Ran}K=K(Z_B^0)$, this means that condition (\ref{s-eq-rel}) is valid if and only if
\begin{equation}\label{supp-cond}
    \varphi_i \cdot S_{y_{d+1},...,y_{s_A}}\varphi_j=0\,\;(\textup{in}\; L_2(\mathbb{R}^{s_A}))\quad
    \forall(y_{d+1},...,y_{s_A})\in\mathbb{R}^{s_A-d},\; \forall i\neq j.
\end{equation}
where $S_{y_{d+1},...,y_{s_A}}$ is a shift operator along the last
$\,s_A-d\,$ coordinates:
$$
(S_{y_{d+1},...,y_{s_A}}\psi)(\xi_1,...,\xi_{s_A})=\psi(\xi_1,...,\xi_d,\xi_{d+1}+y_{d+1},...,\xi_{s_A}+y_{s_A}).
$$
Condition (\ref{supp-cond}) means, roughly speaking, that all shifts
in $\mathbb{R}^{s_A}$ of the supports of all functions of the family
$\{\varphi_i\}$ along the last $\,s_A-d\,$ coordinates do not
intersect each other.
\smallskip

\begin{remark}\label{c-24}
Condition (\ref{supp-cond}) is completely determined by the subspace
$\mathrm{Ran}K$. We will say that \emph{a family $\{\varphi_i\}$ of
functions satisfies condition (\ref{supp-cond}) determined by a
subspace $Z_0$ of $Z_A$} (such that $[Z_0]^{\perp}$ is isotropic) if
it satisfies analog of (\ref{supp-cond}) constructed by using $Z_0$
instead of $\mathrm{Ran}K$. $\square$
\end{remark}\medskip

Thus, in the Schrodinger representation the set
$\Pi(\Phi_{K,\alpha})$ consists of all orthogonal families
$\{\varphi_i\}$ of functions in $L_2(\mathbb{R}^{s_A})$ with unit
norm satisfying condition (\ref{supp-cond}). So, by Theorem
\ref{b-p}A, the equality
$\bar{C}(\Phi_{K,\alpha},\rho)=I(\Phi_{K,\alpha},\rho)$ holds for a
mixed state $\rho$ with finite entropy if and only if this state
$\rho$ has a basis of eigenvectors $\{|\varphi_i\rangle\}$ which (in
the Schrodinger representation) satisfies condition
(\ref{supp-cond}). Theorem \ref{b-p}A also implies that the "if"
part of this assertion is valid for a state $\rho$ with infinite
entropy.\smallskip

By using these observations and by noting that a basis of
eigenvectors of any mixed Gaussian state can not satisfy condition
(\ref{supp-cond}) we obtain the following assertions.
\smallskip

\begin{property}\label{b-l-c} \emph{Let $\,\S_{\mathrm{f}}$ be the subset of $\,\S(\H_A)$ consisting of
all states with finite von Neumann entropy.}\smallskip

A) \emph{If $K\neq0$ (i.e. $\Phi_{K,\alpha}$ is not completely
depolarizing) then
\begin{equation}\label{s-ineq}
 \bar{C}(\Phi_{K,\alpha},\rho)<I(\Phi_{K,\alpha},\rho)
\end{equation}
for any non-degenerate state $\rho\in\S_{\mathrm{f}}$, in particular, for any
non-degenerate Gaussian state $\rho$.}\smallskip

B) \emph{If $\,\mathrm{Ran}K=Z_A$ then (\ref{s-ineq}) holds for any
mixed state $\rho\in\S_{\mathrm{f}}$.}\pagebreak

C) \emph{If $\,[\mathrm{Ran} K]^{\perp}$ is a nontrivial isotropic
subspace of $Z_A$ then
\begin{itemize}
    \item inequality (\ref{s-ineq}) holds for any
mixed Gaussian state $\rho$ and for a mixed state
$\rho\in\S_{\mathrm{f}}$ having no basis of eigenvectors satisfying condition (\ref{supp-cond});
    \item $\bar{C}(\Phi_{K,\alpha},\rho)=I(\Phi_{K,\alpha},\rho)$ for a mixed state $\rho\in\S(\H_A)$
    which has basis of eigenvectors satisfying  condition (\ref{supp-cond}).
\end{itemize}}

D) \emph{If $\,[\mathrm{Ran} K]^{\perp}$ contains a nontrivial
symplectic subspace then there exist mixed Gaussian states $\rho$
such that
$\,\bar{C}(\Phi_{K,\alpha},\rho)=I(\Phi_{K,\alpha},\rho)$.}\footnote{It is precisely case D in which the channel $\Phi_{K,\alpha}$ has completely depolarizing subchannels \cite[Proposition 3]{BRC}. Similar to case C in this case is also possible, by using the Schrodinger representation, to obtain an analogue of condition (\ref{supp-cond}) describing all states $\rho\in\S_{\mathrm{f}}$ for which $\bar{C}(\Phi_{K,\alpha},\rho)=I(\Phi_{K,\alpha},\rho)$.}
\end{property}\medskip

\textbf{Proof.} We have to prove only the last assertion.

If there exists a nontrivial symplectic subspace $Z_{A_0}$ of
$[\mathrm{Ran}K]^{\perp}$ then $Z_A=Z_{A_0}\oplus Z_{A_*}$, where
$Z_{A_*}=[Z_{A_0}]^{\perp}$, and hence
$\H_A=\H_{A_0}\otimes\H_{A_*}$. By using the concatenation rules for
Gaussian channels (see \cite[Ch.12]{H-SCI}) it is easy to show that
$\Phi_{K,\alpha}=\Phi_{K',\alpha}\circ\Psi$, where $\Psi$ is the
partial trace in $\S(\H_A)$ over the space $\H_{A_0}$ and
$\Phi_{K',\alpha}$ is the Gaussian channel from $\S(\H_{A_*})$ to
$\S(\H_{B})$ determined by the "output restriction" $K'$ of the
operator $K$ and the same matrix $\alpha$. It follows that for any
pure Gaussian state $\rho_*=|\psi_*\rangle\langle\psi_*|$ in
$\S(\H_{A_*})$ the subchannel of $\Phi_{K,\alpha}$ corresponding to
the subspace $\H_{A_0}\otimes\,\{\lambda|\psi_*\rangle\}$ is
completely depolarizing. Hence
$\,\bar{C}(\Phi_{K,\alpha},\rho_0\otimes\rho_*)=I(\Phi_{K,\alpha},\rho_0\otimes\rho_*)=0\,$
for any Gaussian state $\rho_0$ in $\S(\H_{A_0})$. $\square$
\medskip

\begin{example}\label{e-1} Applying Proposition \ref{b-l-c} to one-mode Gaussian
channels ($s_A=s_B=1$) we see that  for all such channels excepting
channels of types $A_1$ and $A_2$ (in Holevo's classification
\cite{H-1MGC}) strict inequality (\ref{s-ineq}) holds for all mixed
states with finite entropy.

Type $A_1$ corresponds to completely depolarizing channels. So, channels
of type $A_2$ are the only non-trivial one mode Gaussian
channels for which strict equality (\ref{s-ineq}) is not valid
for all mixed states.

One-mode Gaussian channel of type $A_2$ is a non-discrete c-q
channel, its canonical representative $\Phi_{K,\alpha}$ is
determined by the parameters
$$
K=\left[\begin{array}{cc}
        \;1\;&\; 0\; \\
        \;0\;&\; 0\;
                \end{array}\right],\quad
        \alpha=\left[\begin{array}{cc}
        \;N_0+\textstyle{\frac{1}{2}}\;&\; 0\; \\
        \;0\;&\;N_0+\textstyle{\frac{1}{2}}\;
                \end{array}\right],\;\; N_0\geq 0.
$$
This channel satisfies the condition of part C of Proposition
\ref{b-l-c}. In this case the basis $\{\tilde{e}_k, \tilde{h}_k\}$
introduced in deriving condition (\ref{supp-cond}) consists of the
vectors $\tilde{e}_1=[1,0]^{\mathrm{T}}$, $\tilde{h}_1=[0,1]^{\mathrm{T}}$ and $\mathrm{Ran}K=\{\lambda\tilde{e}_1\}$. So, in the
corresponding Schrodinger representation (in which
$\H_A=\H_B=L_2(\mathbb{R})$) condition (\ref{supp-cond}) is written
as follows
\begin{equation}\label{supp-cond+}
\varphi_i(\xi)\varphi_j(\xi)=0\;\;\textup{almost everywhere
in}\;\,\mathbb{R}\,\;\textup{for all}\;\,i\neq j.
\end{equation}
By Proposition \ref{b-l-c}C,
$\bar{C}(\Phi_{K,\alpha},\rho)=I(\Phi_{K,\alpha},\rho)$ for any
state in
\begin{equation}\label{set}
\!\bigcup_{\{\varphi_i\}\in\Pi(\Phi_{K,\alpha})}\left\{\left.\rho=\sum_i\pi_i|\varphi_i\rangle\langle
\varphi_i|\;\,\right|\,\{\pi_i\}\;\,\textup{is a probability distribution}\,\right\},\!
\end{equation}
where $\Pi(\Phi_{K,\alpha})$ is the set of all families
$\{\varphi_i(\xi)\}$ of functions in $L_2(\mathbb{R})$ with unit
norm satisfying condition (\ref{supp-cond+}). An example of such
family can be constructed by taking a decomposition $\{D_i\}$ of
$\mathbb{R}$ into disjoint measurable subsets and by choosing for
each $i$ a function $\varphi_i(\xi)$ with unit norm vanishing in
$\mathbb{R}\setminus D_i$.

Proposition \ref{b-l-c}C also shows that
$\bar{C}(\Phi_{K,\alpha},\rho)<I(\Phi_{K,\alpha},\rho)$ for any
mixed Gaussian state $\rho$ and for any state $\rho$ with finite
entropy not lying in set (\ref{set}).
\end{example}

\section{On the "complementary" inequality}

It follows from (\ref{b-e-r+}) that conditions for  equality in
(\ref{b-ineq-1}) can be reformulated as conditions for equality in
(\ref{b-ineq-2}) via the notion of a complementary channel. We
focused attention on the former conditions, since they are more
important for applications, while the latter were used in the proofs
as a "bridge" to reversibility properties of a channel.
\smallskip

Theorem \ref{b-p} is reformulated as follows.
\smallskip
\begin{theorem}\label{b-p-comp}
\emph{Let $\,\Phi:\T(\H_A)\rightarrow\T(\H_B)$ be a quantum channel
and $\,\widehat{\Pi}(\Phi)$ the set of all orthogonal families
$\{|i\rangle\}$ of unit vectors in $\H_A$ such that
$\,\supp\Phi(|i\rangle\langle i|)\perp\supp\Phi(|j\rangle\langle
j|)$ for all $\,i\neq j$.}\medskip

\noindent A) \emph{Let $\rho$ be a mixed state in $\S(\H_A)$ such that
$H(\rho)<+\infty$. The following statements are
equivalent:}\smallskip
\begin{enumerate}[(i)]
    \item $\bar{C}(\Phi,\rho)=H(\rho)$;
    \item \emph{the set $\,\widehat{\Pi}(\Phi)$ contains at least one basis of eigenvectors of the state $\rho$;}
    \item \emph{$\Phi|_{\T(\H_{\rho})}$ is isometrically equivalent (see Definition \ref{isom-eq}) to the channel
$$
\varrho\,\mapsto\sum_{i,\,j=1}^{\dim\H_{\rho}}\langle\varphi_i|\varrho\hspace{1pt}|\varphi_j\rangle|\varphi_i\rangle\langle
\varphi_j|
\otimes\sum_{k,\,l=1}^{\dim\H_B}\langle\psi_{jl}|\psi_{ik}\rangle|k\rangle\langle
l|
$$
from $\T(\H_{\rho})$ into $\T(\H_A\otimes\H_B)$, where
$\H_{\rho}=\supp\rho$, $\{|\varphi_i\rangle\}$ is a particular
orthonormal basis of eigenvectors of the state $\rho$,
$\{|\psi_{ik}\rangle\}$ is a collection of vectors in a Hilbert
space such that $\,\sum_{k=1}^{\dim\H_B}\|\psi_{ik}\|^2=1$ for all
$\,i$ and $\{|k\rangle\}$ is an orthonormal basis in $\H_B$.}
\end{enumerate}
\emph{For a state $\rho$ with infinite entropy the above statements
are related as follows
$\mathrm{(ii)\Leftrightarrow(iii)\Rightarrow(i)}$.}\medskip

\noindent B) \emph{The set $\,\widehat{\S}_{\Phi}^{=}$  of all mixed
states $\rho$ in $\S(\H_A)$ with finite entropy, for which
$\mathrm{(i)}$ holds, can be represented as follows
\begin{equation*}
\widehat{\S}_{\Phi}^{=}\;=\bigcup_{\{|i\rangle\}\in\widehat{\Pi}(\Phi)}\left\{\left.\rho=\sum_i\pi_i|i\rangle\langle
i|\,\;\right|\,\{\pi_i\}\in\mathfrak{P}_{\mathrm{f}}\;\right\},
\end{equation*}
where $\mathfrak{P}_{\mathrm{f}}$ is the set of probability
distributions with finite Shannon entropy.}
\end{theorem}\smallskip

\textbf{Proof.} By using the standard representation of a complementary channel (formula (11) in \cite{H-c-c}) and by noting that
$$
\supp\Phi(|i\rangle\langle
i|)\perp\supp\Phi(|j\rangle\langle j|)\quad\Leftrightarrow\quad\widehat{\Phi}(|i\rangle\langle j|)=0
$$
for any vectors $|i\rangle$ and $|j\rangle$ it is easy to show that statements $\mathrm{(ii)}$
and $\mathrm{(iii)}$ in Theorem \ref{b-p-comp} are equivalent respectively to statements $\mathrm{(ii)}$
and $\mathrm{(iii)}$ in Theorem \ref{b-p} for the channel
$\widehat{\Phi}$. $\square$
\smallskip

Since a completely depolarizing channel is complementary to a noiseless (perfectly reversible) channel and vice versa \cite[Ch.10]{H-SCI}, Corollary \ref{b-p+} is reformulated as follows.\smallskip
\begin{corollary}\label{b-p++}
\emph{If $\,\bar{C}(\Phi,\rho)=H(\rho)$ for any 2-rank state $\rho$
then $\,\Phi$ is a noiseless channel, i.e. $\,\Phi$ is unitary
equivalent to the channel $\rho\mapsto\rho\otimes\sigma$, where
$\sigma$ is a given state.}\smallskip
\end{corollary}\medskip

It is well known that the complementary channel to an arbitrary
Gaussian channel is also Gaussian \cite{Caruso,H-SCI}. So, we may
assume that $\Phi_{K,\alpha}=\widehat{\Phi}_{L,\beta}$, where  $L$
is a linear operator $Z_{E}\rightarrow Z_{A}$ and $\beta$ is a real
symmetric $\,(2s_E)\times(2s_E)$ matrix (satisfying the inequality
similar to (\ref{nid})). It follows from Lemma 2 in \cite{BRC} that
\begin{equation}\label{comp-rel}
[\mathrm{Ran} L]^{\perp}=K(\ker \alpha)
\end{equation}
and that the restriction the operator $K$ to the subspace $\,\ker
\alpha\,$ is non-degenerate and symplectic, i.e. it preserves the
corresponding skew-symmetrical forms $\Delta_X$, $X=A,B$. So, we
have
$$
\!\!\begin{array}{c}
\{L=0\}\Leftrightarrow\{\Phi_{K,\alpha}\;\textup{is a noiseless channel}\},\quad
\{\mathrm{Ran}L=Z_A\}\Leftrightarrow\{\det\alpha\neq0\},\medskip\medskip\\
\{\textup{the subspace}\; [\mathrm{Ran} L]^{\perp}\;\, \textrm{is
isotropic}\}\,\Leftrightarrow\,\{\textup{the subspace}\; \ker\alpha\;\,
\textrm{is isotropic}\}.
\end{array}
$$\smallskip

\noindent Hence, by noting that $\,\mathrm{Ran} L=[\mathrm{Ran} L]^{\perp\perp}=[K(\ker \alpha)]^{\perp}$ one can reformulate Proposition \ref{b-l-c} as
follows.\smallskip

\begin{property}\label{b-l-c+} \emph{Let $\,\S_{\mathrm{f}}$ be the subset of $\,\S(\H_A)$ consisting of
all states with finite von Neumann entropy.}\smallskip

A) \emph{If $\,\Phi_{K,\alpha}$ is not a noiseless channel then
\begin{equation}\label{s-ineq+}
 \bar{C}(\Phi_{K,\alpha},\rho)<H(\rho)
\end{equation}
for any non-degenerate state $\rho\in\S_{\mathrm{f}}$, in particular, for any
non-degenerate Gaussian state $\rho$.}\smallskip

B) \emph{If $\;\det\alpha\neq0\,$ then (\ref{s-ineq+}) holds for any
mixed state $\rho\in\S_{\mathrm{f}}$.}\smallskip

C) \emph{If $\;\ker\alpha\,$ is a nontrivial isotropic subspace of
$\,Z_B$ then
\begin{itemize}
    \item inequality (\ref{s-ineq+}) holds for any
mixed Gaussian state $\rho$ and for a mixed state
$\rho\in\S_{\mathrm{f}}$ having no basis of eigenvectors which satisfies condition (\ref{supp-cond}) determined by the subspace
$\,[K(\ker \alpha)]^{\perp}$ (see Remark \ref{c-24});
    \item $\bar{C}(\Phi_{K,\alpha},\rho)=H(\rho)$ for a mixed state $\rho\in\S(\H_A)$ having basis of eigenvectors which satisfies condition (\ref{supp-cond}) determined by the subspace $\,[K(\ker \alpha)]^{\perp}$.
\end{itemize}}

D) \emph{If $\;\ker\alpha\,$ contains a nontrivial symplectic subspace
then there exist mixed Gaussian states $\rho$ such that
$\,\bar{C}(\Phi_{K,\alpha},\rho)=H(\rho)$.}
\end{property}\medskip

\begin{example}\label{e-2} Proposition \ref{b-l-c+}B shows that for all
one-mode Gaussian channels excepting noiseless channels and channels
of type $B_1$ (in Holevo's classification \cite{H-1MGC}) strict
inequality (\ref{s-ineq+}) holds for all mixed states with finite
entropy.

The canonical one-mode Gaussian channel $\Phi_{K,\alpha}$ of type
$B_1$ is determined by the parameters
$$
K=\left[\begin{array}{cc}
        \;1\;&\; 0\; \\
        \;0\;&\; 1\;
                \end{array}\right],\quad
        \alpha=\left[\begin{array}{cc}
        \;\textstyle{\frac{1}{2}}\;&\; 0\; \\
        \;0\;&\;0\;
                \end{array}\right].
$$
In the Schrodinger representation (in which
$\H_A=\H_B=L_2(\mathbb{R})$) condition (\ref{supp-cond}) determined
by the subspace $\,[K(\ker
\alpha)]^{\perp}=\{[\,\lambda,0\,]^{\mathrm{T}}\}$ coincides with
condition (\ref{supp-cond+}).\footnote{This is not surprising and
follows from (\ref{comp-rel}), since the channel of type $A_2$ with $N_0=0$ is
complementary to the channel of type $B_1$ \cite[Ch.12]{H-SCI}.}

Hence, Proposition \ref{b-l-c+}C shows that
$\bar{C}(\Phi_{K,\alpha},\rho)=H(\rho)$ for any state in set
(\ref{set}) and that $\bar{C}(\Phi_{K,\alpha},\rho)<H(\rho)$ for any
mixed Gaussian state $\rho$ and for any state $\rho$ with finite
entropy not lying in set (\ref{set}).
\end{example}\medskip

\section*{Appendix: The proof of Theorem \ref{b-p} in infinite-dimensions}

The proof of the implication $\mathrm{(i)\Rightarrow(iii)}$ in
Theorem \ref{b-p} given in Section 3 is not directly generalized to
the infinite-dimensional case because of possible nonexistence
of an ensemble $\{\pi_i, \rho_i\}$  with average state $\rho$ such
that $\bar{C}(\widehat{\Phi},\rho)=\chi_{\widehat{\Phi}}(\{\pi_i,
\rho_i\})$. This problem can be overcome by using the notion of a
generalized (continuous) ensemble.\smallskip

Following \cite{H-Sh-2} consider an arbitrary Borel probability
measure $\mu$ on the set $\S(\H_A)$ as an input generalized ensemble
for a channel $\Phi:\T(\H_A)\rightarrow\T(\H_B)$ and define its
output $\chi$-quantity as follows
\begin{equation*}
\chi_{\Phi}(\mu)=\int_{\S(\H_A)}H(\Phi(\rho)\Vert\Phi(\bar{\rho}(\mu)))\mu(d\rho)=H(\Phi(\bar{\rho}(\mu)))-\int_{\S(\H_A)}H(\Phi(\rho))\mu
(d\rho),
\end{equation*}
where $\,\bar{\rho}(\mu)=\int_{\S(\H_A)}\rho \mu(d\rho)\,$ is the
barycenter of the measure $\mu$ and the second formula is valid
under the condition $H(\Phi(\bar{\rho}(\mu)))<+\infty$.

Denote by $\P_{p}(\S(\H_A))$ the set of all Borel probability
measures on the set $\S(\H_A)$ supported by pure states. It follows from
Corollary 1 in \cite{H-Sh-2} that
\begin{equation}\label{chi-fun++}
\bar{C}(\Phi,\rho)=\sup_{\bar{\rho}(\mu)=\rho}\chi_{\Phi}(\mu),
\end{equation}
where the supremum is over all measures in $\P_{p}(\S(\H_A))$ with
the barycenter $\rho$.

In contrast to the finite-dimensional case, the supremum in (\ref{chi-fun++}) is not attainable in general,
but there exist sufficient conditions for its attainability, the simplest of them is the following: $H(\Phi(\rho))<+\infty$
\cite[Corollary 2]{H-Sh-2}.\smallskip

Now we are in a position to prove the implication
$\mathrm{(i)\Rightarrow(iii)}$.\smallskip

Assume first that $H(\Phi(\rho))<+\infty$. This and the condition
$H(\rho)<+\infty$ imply $H(\widehat{\Phi}(\rho))<+\infty$ (by the
triangle inequality). By Corollary 2 in \cite{H-Sh-2} there exists a
measure $\mu$ in $\P_{p}(\S(\H_A))$ with the barycenter $\rho$ such
that
$$
\bar{C}(\widehat{\Phi},\rho)=\chi_{\widehat{\Phi}}(\mu).
$$
If $\mathrm{(i)}$ holds then (\ref{b-e-r+}) implies
$\bar{C}(\widehat{\Phi},\rho)=H(\rho)=\chi(\mu)$ and hence
$\chi_{\widehat{\Phi}}(\mu)=\chi(\mu)$. By Proposition 1 in
\cite{HCD} (in which a generalization of Theorem \ref{petz} to
continuous ensembles is established)  this is equivalent to
reversibility of the channel $\widehat{\Phi}$ with respect to a family $\S$ of
pure states such that $\mu(\S)=1$. Let $\S=\bigcup_k\S_k$ be a
decomposition of $\S$ into disjoint non-orthogonally-decomposable
subfamilies and $\{|\varphi_k^i\rangle\}_i$ an orthonormal basis of
eigenvectors of the positive operator
$\rho_k=\int_{\S_k}\rho\mu(\rho)$. Since $\rho=\sum_k\rho_k$ and
$\supp\rho_k\perp\supp\rho_l$ for all $k\neq l$,
$\{|\varphi_k^i\rangle\}_{ik}$ is an orthonormal basis of
eigenvectors of the state $\rho$.

The same arguments as in the finite-dimensional case based on
Theorem \ref{rc} show validity of $\mathrm{(iii)}$ with the basis
$\{|\varphi_k^i\rangle\}_{ik}$.\smallskip

If $H(\Phi(\rho))=+\infty$ then we can choose an increasing sequence
$\{P_n\}$ of finite-rank projectors in $\H_B$ strongly converging to
$I_{\H_B}$ and consider the sequence
$\{\Phi_n\doteq\Pi_n\circ\Phi\}$ of channels from $\T(\H_A)$ into
$\T(\H_B)$, where $\Pi_n(\sigma)=P_n\sigma P_n+[\Tr
(I_{\H_B}-P_n)\sigma]\tau\,$ is a channel from $\T(\H_B)$ into itself, $\tau$
is a given pure state in $\S(\H_B)$. By Lemma \ref{chain}
$\mathrm{(i)}$ implies
$$
\bar{C}(\Phi_n,\rho)=I(\Phi_n,\rho)
$$
for each $n$. Since $H(\Phi_n(\rho))<+\infty$, it follows from the
previous part of the proof that $\mathrm{(iii)}$ and hence
$\mathrm{(ii)}$ hold for the channel $\Phi_n$ for each $n$, i.e.
\begin{equation}\label{ii-exp}
    \Phi_n(|\varphi_i^n\rangle\langle\varphi_j^n|)=0\quad \forall
    i\neq j\;\;\forall n,
\end{equation}
where $\{|\varphi_i^n\rangle\}$ is a  basis of eigenvectors of the
state $\rho$ (depending on $n$).\smallskip

If the state $\rho$ has no multiple eigenvalues then it has the
only (up to permutation and scalar multiplication) basis of
eigenvectors $\{|\varphi_i\rangle\}$ and (\ref{ii-exp}) implies
$$
\Phi(|\varphi_i\rangle\langle\varphi_j|)=\lim_n\Phi_n(|\varphi_i\rangle\langle\varphi_j|)=0\quad
\forall i\neq j,
$$
i.e. validity of $\mathrm{(ii)}$ for the channel $\Phi$.\smallskip

If the state $\rho$ has multiple eigenvalues then the
required basis $\{|\varphi_i\rangle\}$ can be we constructed as follows.

For natural $m$ let $\H_m$ be the direct sum of the eigen-subspaces
of the state $\rho$ corresponding to its $m$ maximal eigenvalues.
Let $d_m=\dim\H_m$. We may consider that the vectors
$|\varphi^n_1\rangle,\ldots|\varphi^n_{d_m}\rangle$ of the above
basis $\{|\varphi^n_i\rangle\}$ belong to the subspace $\H_m$ for
each $n$.\smallskip

Let $n^1_k$ be a sequence of natural numbers such that there exist
$$
\lim_k|\varphi^{n^1_k}_i\rangle=|\varphi^1_i\rangle,\quad
i=\overline{1,d_1},
$$
(existence of this sequence and of the below subsequences follows
from compactness of the unit ball of the subspace $\H_m$ for each
$m$).\smallskip

For $m>1$ let $\,n^m_k$ be a subsequence of $\,n^{m-1}_k$ such that
there exist
$$
\lim_k|\varphi^{n^m_k}_i\rangle=|\varphi^m_i\rangle,\quad
i=\overline{1,d_m}.
$$
It follows from (\ref{ii-exp}) that
\begin{equation}\label{m-eq}
\Phi(|\varphi^m_i\rangle\langle\varphi^m_j|)=\lim_k\Phi_{n^m_k}(|\varphi^{n^m_k}_i\rangle\langle\varphi^{n^m_k}_j|)=0
\end{equation}
for all $i\neq j$ not exceeding $d_m$.\smallskip

By construction $|\varphi^m_i\rangle=|\varphi^{m-1}_i\rangle\,$ for
$\,i=\overline{1,d_{m-1}}$. So, we have the increasing sequence
$$
\{|\varphi^1_i\rangle\}_{i=1}^{d_1}\subset\{|\varphi^2_i\rangle\}_{i=1}^{d_2}\subset\ldots\subset\{|\varphi^m_i\rangle\}_{i=1}^{d_m}\ldots
$$
of orthonormal sets of eigenvectors of the state $\rho$ such that
$\{|\varphi^m_i\rangle\}_{i=1}^{d_m}$ is a basis in the subspace
$\H_m$ for which (\ref{m-eq}) holds.

It is clear that the union $\{|\varphi_i\rangle\}_{i=1}^{+\infty}$
of all the sets of this sequence is a basis of eigenvectors of the
state $\rho$ such that $\Phi(|\varphi_i\rangle\langle\varphi_j|)=0$
for all $i\neq j$. $\square$

\vspace{15pt}

I am grateful to A.S.Holevo and to the participants of his seminar
"Quantum probability, statistic, information" (the Steklov
Mathematical Institute) for useful discussion.

\end{document}